# The optimization for the conceptual design of a 300 MeV proton synchrotron[*]

Yu-Wen An (安宇文)[1,2], Hong-Fei Ji (纪红飞)[1,2], Sheng Wang (王生)[1,2], Liang-Sheng Huang (黄良生)[1,2;1]

[1] Institute of High Energy Physics, Chinese Academy of Sciences (CAS), Beijing 100049, China
[2] Dongguan Neutron Science Center, Dongguan 523803, China

**Abstract:** A research complex for aerospace radiation effects research has been proposed in Harbin Institute of Technology. Its core part is a proton accelerator complex, which consists of a 10 MeV injector, a 300 MeV synchrotron and beam transport lines. The proton beam extracted from the synchrotron is utilized for the radiation effects research. Based on the conceptual design [1], the design study for optimizing the synchrotron has been done. A new lattice design was worked out, and the multi-turn injection and slow extraction system were optimized with the new lattice design. In order to improve the time structure of the extracted beam, a RF knock-out method is employed. To meet the requirement of accurate control of dose, the frequency of the RF kicker is well investigated.

**Key words:** aerospace radiation, slow extraction, RF knock-out method

**PACS:** 41.85.-p, 29.27.Ac

## 1 Introduction

A facility for the simulation of aerospace radiation was proposed and will be constructed in HIT (Harbin Institute of Technology). The accelerator complex, consisting of a 300MeV proton synchrotron and a 10MeV tandem as injector, is the core part of the facility. A conceptual design has been figured out [1]. To further optimize the design of the synchrotron, a new lattice for the synchrotron with many merits has been worked out, and the related issues, injection and extraction, have been further investigated and optimized.

The synchrotron is designed to accumulate and accelerate the proton beam from 10 MeV to 300 MeV. The beam energy of the research complex could be adjusted continuously from 10 MeV to 300 MeV with maximum repetition rate of 0.5 Hz. The extraction beam is used to do research for material, electronic component and biology, so the stable micro beam and the scanning beam should be available. The main parameters of the synchrotron are listed in Table 1 [1]. The horizontal acceptance of the synchrotron is 180 $\pi$ mm-mrad, and the momentum deviation is 3E-3 in injection stage.

Table 1: Main parameters of the synchrotron of the aerospace radiation facility

| Parameters | Units | Values |
|---|---|---|
| Ring Circumference | M | 35 |
| Inj. Energy | MeV | 10 |
| Ext. Energy | MeV | 300 |
| Nominal Tunes(H/V) |  | 1.72/1.28 |
| Natural Chromaticity |  | -1.25/-0.68 |
| Repetition Rate | Hz | 0.5 |
| Horizontal Acceptance | $\pi$ mm-mrad | 180 |
| Accumulated Proton Number |  | $2*10^{11}$ |
| Injection Momentum Deviation |  | $3*10^{-3}$ |
| Harmonic Number |  | 1 |

In Sec. 2, we describe briefly the new lattice design of the synchrotron of the aerospace radiation complex. Sec. 3 is devoted to the design of injection system of the synchrotron. The design of extraction system is shown in Sec.4, and in order to achieve a low ripple of the extraction beam, the RF knock-out method is employed and the parameters of RF kicker are carefully optimised. The physical aperture and dynamic aperture are discussed in Sec. 5. A short summary is presented in Sec.6.

[*] Supported by National Natural Science Foundation of China (11405189)
1) Corresponding author (email: huangls@ihep.ac.cn)

## 2 New Lattice design of the synchrotron

Based on the original parameters on the conceptual design [1], a compact new lattice was worked out for synchrotron. The short circumference will induce the high field of magnets, and too much high field may increase the manufacturing difficulties and cost of magnets. The circumference of 35m was carefully chosen for a compromised consideration. The layout of the synchrotron is shown in fig. 1. The arc of the synchrotron is composed of FODO cells, and the two achromatic straight sections of 8 m that connect arcs are reserved for the elements of injection and extraction, RF cavities and resonance sextupoles. 8 dipoles and 12 quadrupoles are distributed separately along the ring, and the maximum field of dipoles is 1.51 T. Fig. 2 [2] shows the twiss parameters of the synchrotron. Short drifts in the arc with large dispersion are reserved for correctors, BPMs, chromatic sextupoles, RF kicker and so on.

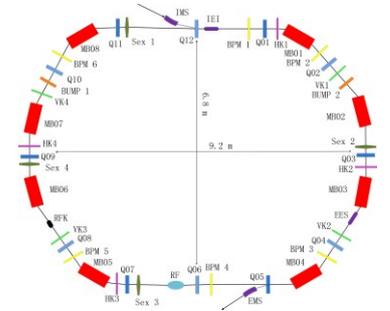

**Fig. 1.** The layout of the synchrotron. The IEI (injection electric inflector) and the IMS (injection magnet septum) are septa for injection, and the EES (extraction electric septum) and EMS (extraction magnet septum) are septa for extraction.

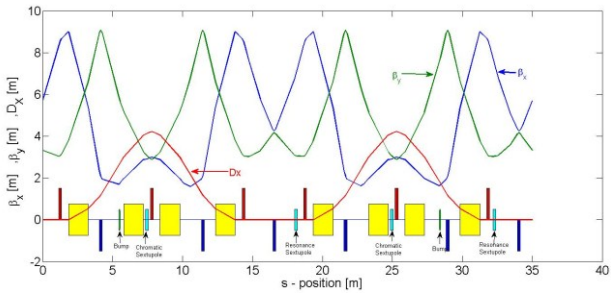

**Fig. 2.** The twiss parameters of the synchrotron of the aerospace radiation facility. The horizontal betatron function, vertical betatron function and horizontal dispersion function are represented with blue line, green line and red line respectively.

The horizontal betatron function and vertical betatron function are designed to be smaller than 10 m, and the horizontal dispersion function is smaller than 4 m as well, and the advantage of this design is that the larger dispersion era along with smaller beta function do not need extra aperture of the magnets. Although the natural chromaticity, arising solely from quadrupoles, is not very large in our design, the chromaticity need to be corrected to meet the demand of the low ripple of the extraction beam.

The closed orbit distortion (COD) is mainly from the field errors and misalignment of dipoles and quadrupoles. In the process of injection, larger COD can severely reduce the injection efficiency for the decreasing acceptance of the ring. Besides, larger COD also severely reduces the extraction efficiency for a bad spill structure of the beam during the slow extraction period. In order to correct COD distortion, 6 bipolar BPMs (BPM1~BPM6 as shown in Fig. 1), 4 vertical steering magnets (VK1~VK4 as shown in Fig. 1) and 4 horizontal steering magnets (HK1~HK4 as shown in Fig. 1) are reserved to correct COD toward less than 1 mm.

## 3 Injection system design

A tandem was chosen as the injector of the synchrotron, and it can provide 10MeV proton beam. The multi-turn injection scheme is suitable for accumulating protons in the synchrotron. Electromagnetic septa are the devices with a thin structure of coils or electrodes [3]. Two types of the septa, magnetic septa and electric septa are often used for multi-turn beam injection, and the magnetic septa are often used to tilt a large angle while electric septa tilt a small angle. For taking multi-turn injection scheme, a bump generated by two dipoles with phase advance of $\pi$ locates nearly symmetrically along the ring. In the process of injection simulation, the emittance of painting beam is 80 $\pi$mm-mrad. Fig. 3 shows the variation of the orbit of the circulating beam during injection, in which the green line represents the injection point. The power supply of the bump magnets is ramped down after 20 turns injection, and the designed orbit of the circulating beam jumps down to zero.

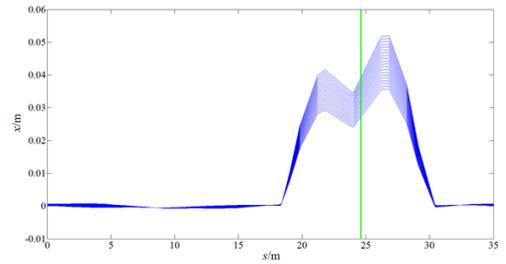

**Fig. 3.** The orbit variation of the circulating beam during injection and the green line represents the injection point.

In our design, the proton beam from the injector is tilted by the injector magnetic septum (IMS), and then goes through quadrupole Q12 and tilted with a small angle by the injection electric inflector (IEI) and converged to the circulating beam. The height of the IEI is 48 mm and the particles distribution in phase space at the end of 20 turns injection is shown in Fig. 4. The particles accumulated in the synchrotron w.r.t injection number is shown in Fig. 5, and 1000 macro particles per turn are injected into the ring.

The injection efficiency, defined as ratio of survival particles and injected particles, is around 40%, and the beam is mainly lost at IEI.

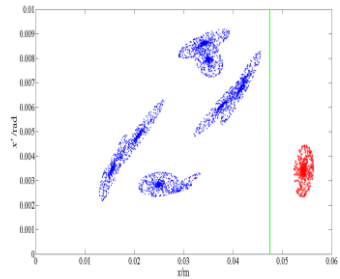

**Fig. 4.** The particles distribution in phase space at the end of 20 turns injection, and the green line represents of the location of the IEI.

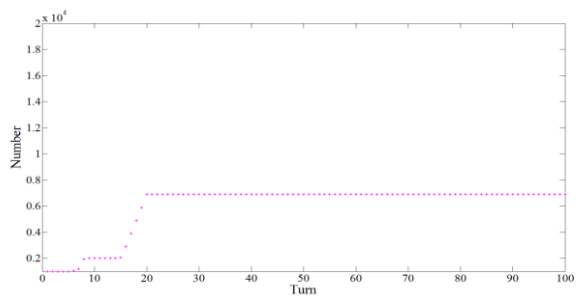

**Fig. 5.** The particles accumulated in the synchrotron w.r.t injection turn.

## 4 Extraction system design

### a) Ordinary slow extraction pattern

Either half integer or third-order integer resonance can be used for slow extraction, but the current trend is towards using third-order resonance for more controllable spills [4]. It is difficult to use fourth order resonance because the separatrices are becoming too small [5]. As shown in Fig. 1, the extraction electric septum (EES) and the extraction magnet septum (EMS) are the septa for slow extraction, and sextupole 1 and sextupole 3 are resonance sextupoles while sextupole 2 and sextupole 4 are chromatic sextupoles. The phase space is shaped into triangle by sextupoles, particles inside the triangle are stable while particles out of the triangle are unstable. In the ordinary slow extraction scheme, the size of the triangle can be shrunk by adjusting the resonance sextupoles or tuning the horizontal tune. The ordinary slow extraction method is related to the slow beam responding, and tuning quadrupoles or sextupoles can make the disturbance of lattice symmetry. So the spill structure of the ordinary slow extraction is not good for special radiation application for its requirement of irregular dose. Fig. 6 shows the spill structure of the extracted beam with resonance sextupoles obeys the Rayleigh distrition, and the corresponding momentum deviation is 0.1%.

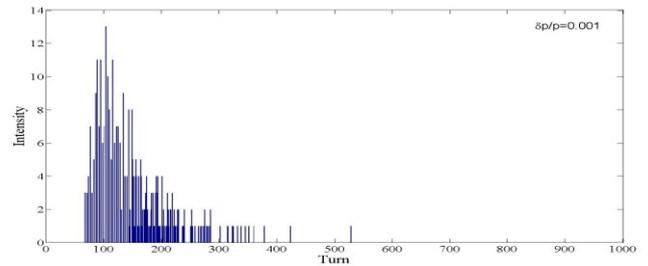

**Fig. 6.** The spill structure of the extracted beam only with resonance sextupoles.

### b) Slow extraction by RF knock-out method

To improve time structure of the extracted beam, RF knock-out method is often employed in the slow extraction application. RF knock-out method is a good option because the parameters of the lattice and spiral step of extracted particles can keep constant during the slow extraction period. Besides, RF knock-out method can provide high irradiation accuracy even for an irregular target, and results in high beam-utilization efficiency. The RF knock-out method is also useful in the spot scanning method, because the beam supply can be easily started or stopped in the extraction period [6]. RF kicker is an accelerator dipole that can be adiabatically turned on to induce coherent betatron oscillation without causing serious emittance dilution [7]. The ratio of the RF kicker frequency and the beam revolution frequency is related to the fractional part of the horizontal tune of the synchrotron. However, in order to cover the tune spread of the revolution particles, the frequency of the RF kicker is always set with a bandwidth $\Delta f_k$ [8], so the frequency of the RF kicker is a periodic saw-tooth wave function, and the repetition cycle of RF kicker frequency modulation is called a FM period.

In our simulation, the central frequency of the RF kicker is 3.746 MHz with the bandwidth of 20 KHz, and the maximum kick angle of the RF kicker is 5 $\mu rad$. The repetition cycle of the frequency modulation is set to 1 ms, and the energy of the proton beam is 300 MeV. The tune of the synchrotron is 1.67/1.21 (H/V), and the horizontal natural chromaticity is corrected to 0.2 by the chromatic sextupoles. The spill structure of the extracted beam is shown in Fig. 7. And we can observe that the spill structure of the extracted beam with a proper intensity has no-beam periods in the extraction duration.

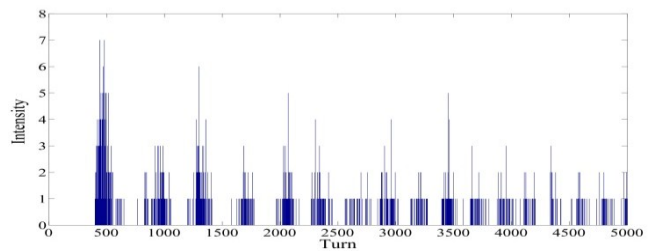

**Fig. 7.** The spill structure of the extracted beam with RF kicker.

### c) Slow extraction by RF kicker with dual Frequency modulation method

For improving the irradiation accuracy, many advance techniques have been proposed and well applied in HIMAC (Heavy Ion Medical Accelerator in Chiba) [9]. One of these techniques is called dual FM (Frequency Modulation) method. Compared with ordinary RF kicker knock out method (single FM), dual FM method can modulate the RF kicker frequency more robust, and that is very useful and helpful to extract core particles in the central part of the separatrix. The kick angle of RF kicker with dual frequency modulation can be expressed as

$$\theta_{rf-kicker} = \theta_{max}\left\{\sin\int_0^t[2\pi f_k + g_1(\tau)]d\tau + \sin\int_0^t[2\pi f_k + g_2(\tau)]d\tau\right\}, \quad (1)$$

where $g_1(\tau)$ and $g_2(\tau)$ are related to the frequency bandwidths of the two electrodes of the RF kicker, and $f_k$ is the central frequency of the RF kicker. In our simulation, the maximum kick angle of the two RF kicker electrodes is 3 $\mu rad$, and the frequency bandwidth of the two RF kicker electrodes is shown in Fig. 8.

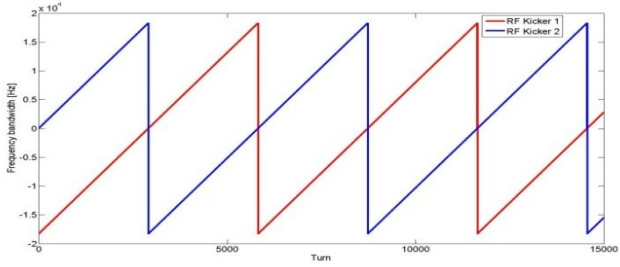

**Fig. 8.** The frequency bandwidths of RF kicker with dual FM. The red saw-tooth wave function represents the first RF kicker electrode frequency bandwidth and the blue saw-tooth wave function represents the second RF kicker electrode frequency bandwidth respectively.

As described in single FM section, RF knock-out method with single FM has the disadvantage that the no-beam period exists in one FM period, that because this technique can only extract particles in the extraction regions [9]. The extraction regions and the diffusion regions are two regions inside of the separatrix, and the main difference between them is the different responding to the frequency of the RF kicker. Compared with RF kicker with single FM, dual FM can extract beam of the no-beam period in one FM period. That is, particles located in the diffusion regions can be shifted to extraction region with the growth of the particles amplitude caused by the other RF kicker electrode, and these particles can be extracted in no-beam period corresponding to the RF kicker with single FM. Fig. 9 shows the spill ripple of the extracted beam with dual FM RF knock-out method, and the spill structure is well made more uniform compared with that shown in Fig. 7.

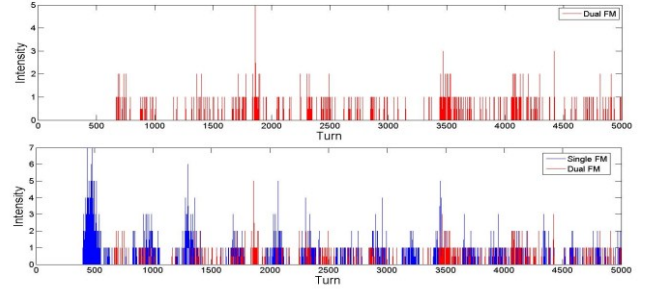

**Fig. 9.** Bottom: The comparison of spill structure between single FM and dual FM. Top: The spill structure of the extracted beam with dual FM.

## 5 Physical aperture and dynamic aperture

Considering the nonlinear effects come from sextupoles and higher order field of quadrupoles and dipoles, the dynamic aperture of the synchrotron was studied. The dynamic aperture of the synchrotron at the injection point is shown in Fig. 10, and the different color lines represent different momentum deviations. For the designed horizontal acceptance 180 $\pi$mm-mrad, the dynamic aperture is much larger than the physical aperture.

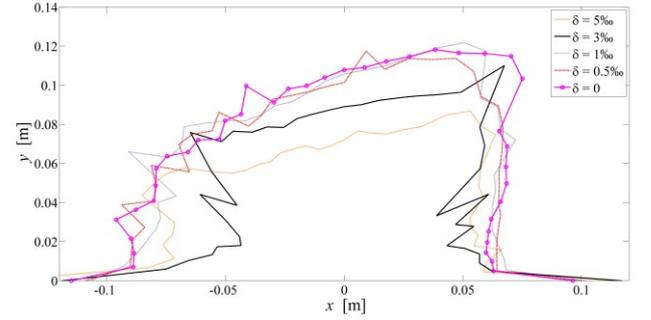

**Fig. 10.** The dynamic aperture of the synchrotron. Different color lines correspond to different momentum deviations.

## 6 Summary

The synchrotron of the aerospace radiation complex with circumference of 35 m has been designed. The design of the new lattice, injection system, and extraction system is described in this paper, and it is verified by the simulation. For high injection efficiency, the decreasing pattern of the bump and is carefully optimised. For a flatten spill structure of the extracted beam, RF kicker with frequency modulation is employed in our design. Owing to improving the spill structure, an advanced technique called dual frequency modulation is imported, and the spill structure is well suppressed. The dynamic aperture of the synchrotron was studied and discussed in the paper.

*We wish to thank Zhang Manzhou in SSRF (Shanghai Synchrotron Radiation Facility) for his helpful and useful discussions on third order resonance slow extraction.*